\documentclass{jetpl}
\twocolumn

\lat

\title{A doublet of cosmic-ray events with primary energies $>10^{20}$~eV}

\rtitle{A doublet of cosmic-ray events\ldots}

\sodtitle{A doublet of cosmic-ray events with primary energies
$>10^{20}$~eV}

\author{S.\,V.\,Troitsky$^{+}$\thanks{e-mail: st@ms2.inr.ac.ru}}

\rauthor{S.\,V.\,Troitsky}

\sodauthor{S.\,V.\,Troitsky}

\address{$^+$Institute for Nuclear Research of the Russian Academy of
Sciences,\\
60th October Anniversary prospect 7A, 117312 Moscow, Russia}

\dates{May 29, 2012}{*}

\abstract{
The Telescope Array Collaboration has observed a cosmic-ray event with
estimated primary energy of $1.38 \times 10^{20}$~eV whose arrival
direction coincides \cite{TA-anisotropy}, given the angular resolution of
$1.5^{\circ}$, with that of an event with estimated primary energy of
$1.23 \times 10^{20}$~eV observed by the Pierre Auger Observatory. The
total number of events with energies $>10^{20}$~eV published by both
experiments is six. I estimate the statistical significance of the
doublet, which is rather weak, and point out that the arrival directions
of events in the doublet coincide with the Galactic X-ray source Aql~X-1.}

\PACS{98.70.Sa}

\begin{document}

\maketitle

Despite decades of intense studies, including those by recent huge
experiments, sources of ultra-high-energy cosmic rays (UHECRs) remain
unknown. For the primary cosmic-ray particles with energies of order
$10^{20}$~eV or higher, quite simple astrophysical arguments restrict the
number of potential accelerating astrophysical sources drastically
\cite{Hillas, ourHillas}. At the same time, the Greizen \cite{G}, Zatsepin
and Kuzmin \cite{ZK} (GZK) effect shortens the mean free path of protons
and nuclei with those high energies considerably, putting an additional
constraint that sources of these events should be nearby. This logic has
been supported by the recent observations of the flux suppression at high
energies consistent with the GZK predictions by the High-Resolution Fly's
Eye (HiRes) \cite{HiRes-GZK}, Pierre Auger Observatory (PAO)
\cite{Auger-GZK} and Telescope Array (TA) \cite{TA-GZK} experiments. The
observation of the suppression does not mean however that no ``super-GZK''
events are observed. Two largest and most modern arrays of surface
detectors have published coordinates of three events each
\cite{PAO-events, TA-anisotropy} with energies $E>10^{20}$~eV. It is these
six events which will be primarily concerned in this note.

For the ``sub-GZK'' ($E\sim 6 \times 10^{19}$~eV) events, early
PAO data suggested a weak correlation of cosmic-ray arrival
directions with positions of nearby active galactic nuclei (AGN)
\cite{PAO-Science} which might be interpreted as an indication to
acceleration of UHECRs in these astrophysical objects. This interpretation
has been criticised on the basis of numerous arguments, see e.g.\
\cite{Comment, Moscalenko, Gureev}; it has not been supported by the data
of HiRes \cite{HiRes-AGN} and TA ~\cite{TA-anisotropy} though it has
been supported by the Yakutsk data \cite{Yakutsk-AGN}. A subsequent
publication of the Pierre Auger collaboration \cite{PAO-events}, based on
enlarged statistics, demonstrated a much weaker effect. However,
the highest-energy
events with $E>10^{20}$~eV did not correlate with AGN even in the data set
with the strongest signal.

\begin{figure}
\includegraphics[width=0.95\columnwidth]{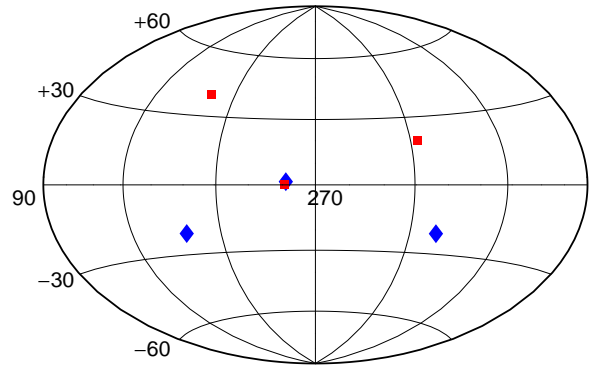}
\caption{\label{fig:skymap}
Figure~\ref{fig:skymap}. The sky map with arrival directions of three
PAO events with $E>10^{20}$~eV (diamonds) and three TA events with
$E>10^{20}$~eV (boxes). The Hammer projection, equatorial coordinates.}
\end{figure}

\begin{figure}
\includegraphics[width=0.95\columnwidth]{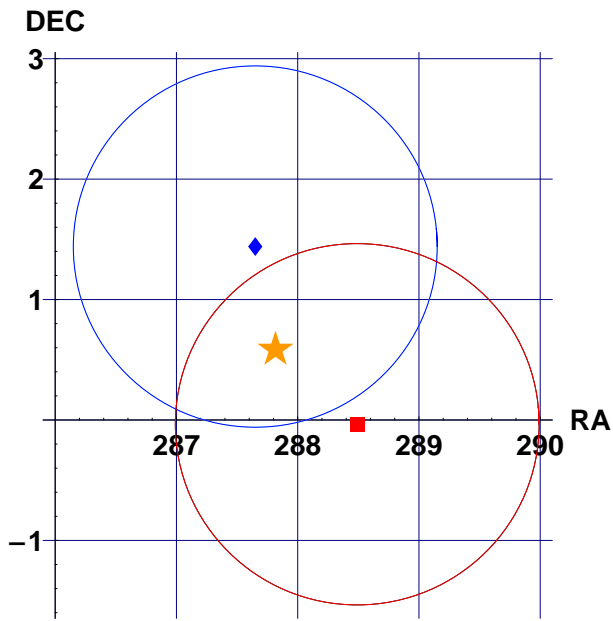}
\caption{\label{fig:zoom}
Figure~\ref{fig:zoom}. The sky map with arrival directions of the two
events in the doublet: the PAO event (diamond) and
the TA event (box). With a 68\% probability, the true arrival directions
are inside the corresponding circles. The star denotes the position of
Aql~X-1; no other strong X-ray or gamma-ray sources are seen nearby.}
\end{figure}

As it has been pointed out in recent Ref.~\cite{TA-anisotropy}, where
coordinates of TA events have been made public for the first
time, the arrival direction of one TA event with $E>10^{20}$~eV coincides,
within the experimental precision, with that of a PAO event of the similar
energy. Details of the two events are given in Table~\ref{tab:doublet} for
convenience.
\begin{table}
\begin{center}
\begin{tabular}{ccccc}
\hline
Exp. & Date & $E$, EeV & RA & DEC \\
\hline
PAO  & 09.10.2008 & 123 & 287.7$^{\circ}$ & $+1.4^{\circ}$\\
TA  & 28.02.2011 & 138 & 288.5$^{\circ}$ & $-0.0^{\circ}$\\
\hline
\end{tabular}
\caption{\label{tab:doublet} Table~\ref{tab:doublet}.
Details of the two events with coinciding arrival
directions: the experiment name; date; energy in units of
EeV$=10^{18}$~eV; equatorial coordinates. }
\end{center}
\end{table}
The sky map with all six events with $E>10^{20}$~eV is presented in
Fig.~\ref{fig:skymap}.
The appearence of the doublet is psychologically surprising because the
two experiments are located in different hemispheres and see different
parts of the sky with a moderate overlap in the equatorial region.
The zoom of the skymap, with error circles of events, is presented in
Fig.~\ref{fig:zoom}.

To estimate the statistical significance of this doublet we follow the
usual procedure described in Ref.~\cite{TT:auto} (see also
\cite{TT:penalties, Finley}). We assume the isotropic distribution of
arrival directions, account for direction-dependent experimental exposure
and simulate a sufficient number of Monte-Carlo sets of events, then count
how often the same or larger number of doublets happens as a fluctuation
of the isotropic distribution. For this purpose, a doublet is defined as a
pair of arrival directions separated by not more that $\sqrt{2}
\theta_{0}$ where the angular resolution $\theta_{0}\approx 1.5^{\circ}$
for both PAO and TA.

For three PAO and three TA events ($E>10^{20}$~eV), the $P$-value
calculated in this way is $P\approx 3.7 \times 10^{-3}$. This value may be
interpreted as an estimate of the probability to have one doublet anywhere
in the combined data set as a fluctuation of the isotropic distribution of
arrival directions.
However, this interpretation should be taken with
care because the choice of the energy threshold, $10^{20}$~eV, is somewhat
arbitrary.

We see that the statistical significance of the doublet is not that
impressive. Many three-sigma effects have come and gone in cosmic-ray
physics. Nevertheless, it is tempting to speculate about the potential
source of the particles. The region of interest is located close to the
Galactic plane, in the zone of avoidance where not many extragalactic
objects are identified. However, strong active galaxies which might
accelerate particles up to ultra-high energies are expected to be X-ray
and/or gamma-ray sources visible through the dust obscuration at this
location (Galactic coordinates of the center of the doublet are $l\approx
35.9^{\circ}$, $b\approx -4.3^{\circ}$). In representative catalogs
of active galactic nuclei (Veron-Cetti and Veron \cite{Veron}), gamma-ray
sources (2FGL \cite{2FGL}) and X-ray sources (ROSAT bright source catalog
\cite{ROSAT}), there is only one bright source within a few degrees of
this location, a low-mass X-ray binary star Aql~X-1, just in the middle of
the doublet (see Fig.~\ref{fig:zoom}). There are no active galaxies nor
other identified ROSAT X-ray or FERMI-LAT gamma-ray sources around.

\begin{figure}
\includegraphics[width=0.95\columnwidth]{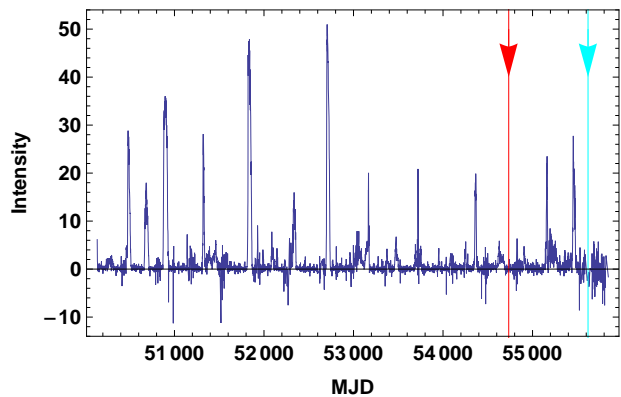}
\caption{\label{fig:lightcurve}
Figure~\ref{fig:lightcurve}. The X-ray light curve of Aql~X-1
(quick-look results provided by the ASM/RXTE team \cite{RXTE}). Vertical
lines with arrows denote the arrival times of cosmic rays. }
\end{figure}

Aql~X-1, the brightest X-ray source in the Aquilla constellation, is an
X-ray millisecond pulsar in a binary system (see e.g.\
Ref.~\cite{Aql-recent} for discussion and references). The system is
located at the distance of $5.2^{+0.7}_{-0.8}$~kpc from the Earth
\cite{Aql-dist}. It experiences quasi-periodic outbursts each 300 days
roughly (see the X-ray light curve in Fig.~\ref{fig:lightcurve}).
Though the object is one of only twelve known Galactic
accretion-powered millisecond pulsars \cite{9psrs} and is well studied, it
does not appear very exotic. It is singled out of this dosen only by a
relatively large mass of the companion in the binary system, $M\gtrsim
0.45 M_{\odot}$, and the correspondingly large orbital period of $\sim
19$~h. The estimated magnetic field on the neutron-star surface is $\sim
(1\dots 5)\times 10^{8}$~G \cite{Aql-MF1, Aql-MF2}. On the basis of
X-ray timing and spectral properties, this object is classified as
``atoll'' (see e.g.\ Ref.~\cite{NSreview} for a more detailed discussion
of classifications). Accretion in these sources may have similarities with
accretion on stellar-mass black holes \cite{AtollSimBH}.

An extragalactic $E\sim 10^{20}$~eV proton arriving
from the direction we consider should be deflected by the magnetic field
of the Milky Way by $\sim (2\dots 4)^{\circ}$, depending on the field
model which is not known precisely. However, a hypothetical particle
coming from Aql~X-1 would be deflected by a much smaller amount because the
source is considerably closer to the Earth than the Galcatic Center is.
Assuming charge one and the mean Galactic magnetic field in the disk of
$\sim 1~\mu$G, one obtains a rough estimate for the deflection of $\sim
1.1^{\circ}$. This deflection would correspond, for a proton, to the time
delay of $\sim $~yr, thus making it not surprising that the arrival
times of the events do not coincide with  particularly interesting moments
in the life of the would-be source, cf.\ Fig.~\ref{fig:lightcurve} (unless
neutral primaries are assumed). We note that
currently, neither PAO nor TA is able to determine the primary particle
type of a particular air shower detected by the surface array.

In general, a wide belief that cosmic rays with $E \gtrsim 10^{19}$~eV are
of extragalactic origin is based on a few reasonable arguments. First,
these energetic particles are not confined by the Galactic magnetic field
and, assuming similar fields exist in other galaxies, are not confined
anywhere. Second, the arrival directions of these cosmic rays are (almost)
isotropic on large angular scales, while the distribution of any kind of
Galactic sources on the sky is anisotropic. Third, there is a lack of
Galactic objects where sufficient conditions for acceleration of
particles to those energies are satisfied. Nevertheless, some proposals
for Galactic sources are being discussed (see e.g.\ Refs.~\cite{Kus-gal,
Far-gal, Oli-gal}).

The first two arguments might be overcome if in
the Milky Way there are only a few sources of cosmic rays. Then, part of
the observed events come from these few Galactic sources and the rest
comes from similar sources in nearby galaxies (thus explaining weak hints
of correlation with the local distribution of matter at the highest
energies). However, Galactic sources should then dominate the flux at
high energies (cf.\ e.g.\ Ref.~\cite{Dubovsky}) thus producing either the
Galactic anisotropy or a concentration of arrival directions towards
particular sources. A price to pay for not seeing this in data is the fine
tuning which is however not excluded.

As for the third argument, acceleration of UHECRs in pulsars has been
proposed long ago in Ref.~\cite{PSR-old} (see Ref.~\cite{Oli-gal} for a
different recent proposal). In general, it is difficult to overcome
radiative energy losses in pulsar magnetosphere~\cite{ourHillas,
Derishev};
however, in the regime where the losses are dominated by the curvature
radiation and the electric and magneic fields are parallel in a very long
tube (``linear accelerator''), the energy-loss limits may be relaxed
\cite{ourHillas, Derishev}. Another problem with pulsars is the screening
of the accelerating potential gap when electron-positron pairs are
created; however, one expects that some pulsars are ``pair-starved''
\cite{pair-starved0, pair-starved} and may accelerate particles to higher
energies. It is presently unclear whether these conditions are satisfied
in Aql~X-1.

To summarize, we observe an interesting coincidence of the arrival
directions of two out of six cosmic-ray particles with estimated energies
in excess of $10^{20}$~eV observed by the modern surface-detector arrays,
the Pierre Auger Observatory and the Telescope Array experiment. The
probability that one or more doublets occur in the data set by a
fluctuation of the isotropic flux is about $3\times 10^{-3}$. The error
circle of the arrival directions of coinciding cosmic rays includes a
bright accretion-powered millisecond pulsar in the X-ray binary Aql~X-1.
Future studies are required to support or reject the conjecture that
Galactic sources are able to produce super-GZK cosmic rays and whether
this particular source possesses physical conditions allowing for particle
acceleration to that high energy.

I acknowledge interesting discussions with my colleagues from the Telescope
Array experiment as well as with S.~Popov and M.~Revnivtsev.
This work was supported in part by the RFBR grants 10-02-01406 and
11-02-01528, by the grant of the President of the Russian Federation
NS-5590.2012.2 and by the ``Dynasty'' foundation.

\end{document}